\documentclass[12pt]{article}
\usepackage{amsmath,amssymb,graphicx} 
\usepackage{epsf}
\usepackage{epsfig}
\usepackage{pstricks}
\usepackage{axodraw}
\usepackage{cite}

\newcommand{\met}{\not\!\!\!E_T}

\newcommand{\be}{\begin{equation}}
\newcommand{\ee}{\end{equation}}

\newcommand{\beq}{\begin{eqnarray}}
\newcommand{\eeq}{\end{eqnarray}}

\newcommand{\centeron}[2]{{\setbox0=\hbox{#1}\setbox1=\hbox{#2}\ifdim
\wd1>\wd0\kern.5\wd1\kern-.5\wd0\fi
\copy0

\kern-.5\wd0\kern-.5\wd1\copy1\ifdim\wd0>\wd1
                                       \kern.5\wd0\kern-.5\wd1\fi}}
\newcommand{\ltap}{\>\centeron{\raise.35ex\hbox{$<$}}
                               {\lower.65ex\hbox{$\sim$}}\>}
\newcommand{\gtap}{\>\centeron{\raise.35ex\hbox{$>$}}
                               {\lower.65ex\hbox{$\sim$}}\>}

\newcommand\ZZ{\hbox{\zfont Z\kern-.4emZ}}
\font\zfont = cmss10 

\begin{document}
\begin{titlepage}
\begin{flushright}
\end{flushright}

\vskip.5cm
\begin{center}
{\huge \bf 
Discovering hidden sectors with mono-photon $Z'$ searches
}

\vskip.1cm
\end{center}
\vskip0.2cm

\begin{center}

{\bf   Yuri Gershtein}
\vskip 8pt
{\it Department of Physics and Astronomy, Rutgers University, Piscataway, NJ 08854} \\
\vskip 8pt
{\bf   Frank Petriello, Seth Quackenbush}
\vskip 8pt
{\it Department of Physics, University of Wisconsin, Madison, WI 53706, USA} \\
\vskip 8pt
{\bf  Kathryn M. Zurek}
\vskip 8pt
{\it Department of Physics, University of Wisconsin, Madison, WI 53706, USA \\ and \\ Center for Particle Astrophysics, Fermi National Accelerator Laboratory, Batavia, IL 60510} \\
\end{center}

\begin{abstract}
\vskip 3pt
\noindent
In many theories of physics beyond the Standard Model, from extra dimensions to Hidden Valleys and models of dark matter, $Z'$ bosons mediate between Standard Model particles and hidden sector states.  We study the feasibility of observing such hidden states through an invisibly decaying $Z'$ at the LHC.  We focus on the process $pp \rightarrow \gamma Z' \rightarrow \gamma X X^\dagger$, where $X$ is any neutral, (quasi-) stable particle, whether a Standard Model (SM) neutrino or a new state.  This complements a previous study using $pp \rightarrow Z Z' \rightarrow \ell^+ \ell^- X X^\dagger$.  Only the $Z'$ mass and two effective charges  are needed to describe this process.   If the $Z'$ decays invisibly only to Standard Model neutrinos, then these charges are predicted by observation of the $Z'$ through the Drell-Yan process, allowing discrimination between $Z'$ decays to SM $\nu$'s and invisible decays to new states.  
We carefully discuss all backgrounds and systematic errors that affect this search.  We find that 
hidden sector decays of a 1 TeV $Z'$ can be observed at $5\sigma$ significance with $50\,{\rm fb}^{-1}$ at the LHC.  Observation of a 1.5 TeV state requires super-LHC statistics of $1\,{\rm ab}^{-1}$.  Control of the systematic errors, in particular the parton distribution function uncertainty of the dominant $Z\gamma$ background, is crucial to maximize the LHC search 
reach.

\end{abstract}

\end{titlepage}

\newpage


\section{Introduction}
\label{sec:intro}

New massive $U(1)$ gauge bosons appear in numerous theories of physics beyond the Standard Model (SM).  They appear in grand unified theories such as $SO(10)$~\cite{mohapatra} 
and $E(6)$~\cite{Hewett:1988xc}, in theories of extra space-time dimensions as Kaluza-Klein excitations of the SM gauge bosons~\cite{Hewett:2002hv}, and in 
Little Higgs theories of the 
electroweak sector~\cite{Schmaltz:2005ky}.  $Z'$ bosons that decay to leptons have a simple, clean experimental signature, and consequently can be searched for up to high 
masses at colliders.  Current direct search limits from Tevatron experiments restrict the $Z'$ mass to be greater than about 900 GeV when its couplings to SM fermions are identical to those of the $Z$ boson~\cite{Tev:2007sb}.  The LHC experiments are expected to extend the $Z'$ mass reach to more than 5 TeV~\cite{tdrs}.

$Z'$ bosons often serve as messengers which connect the SM to hidden states, such as in some models of supersymmetry breaking~\cite{Chung:2003fi}, extra dimensions \cite{Dobrescu}, and in Hidden Valley models~\cite{Strassler:2006im}.  
The $Z'$ can decay to light particles in these hidden sectors.  
Hidden Valley models, for example, contain sub-TeV mass states which are electrically neutral and quasi-stable, with decay lengths in some cases longer than tens of meters.  These exit the detector as missing energy.  A sterile neutrino which is charged under the $U(1)'$ would also result in hidden decays of the $Z'$.  In certain extensions of the MSSM, invisible decays of the $Z'$ are predicted \cite{Pierce}.  In models of extra dimensions, the $Z'$ may decay invisibly to Kaluza-Klein neutrinos \cite{Dobrescu,Schmaltz}.  Such states may also account for the observed dark matter, as in the model of Ref.~\cite{Hooper:2008im}; a model of milli-charged dark matter from a Stueckelberg $Z'$ may be found in Ref.~\cite{Cheung}.  
Analysis of $Z'$ bosons decaying to hidden states is complimentary to studies where instead the Higgs boson acts as the messenger to a hidden sector\cite{Wells,Strassler}; the phenomenology of such a scenario was studied in \cite{Han}.

In this paper we study invisible decays of $Z'$ bosons, and consider whether such decays can be detected at the LHC using the mono-photon channel $pp \to \gamma Z' \to \gamma \met$.  Our study 
extends a previous study of the  $pp \to ZZ' \to \ell^+ \ell^- \met$ mode~\cite{Petriello:2008pu}.  The mono-photon signature has a simpler structure than the 
$ZZ'$ mode.  Only three parameters describe the process: the $Z'$ mass and two effective charges associated with the $Z'$ couplings to quarks and hidden states.  If the $Z'$ decays leptonically, the charges are predicted by on-peak measurements in the Drell-Yan channel assuming invisible $Z'$ decays to only neutrinos~\cite{Petriello:2008zr}. 
This allows the presence of hidden states coupled to the $Z'$ to be probed.  If the $Z'$ is leptophobic, then this becomes a discovery mode.  We illustrate how to separate the $Z'$ signal from background using as an example 
the $U(1)_{\chi}$ model with a vector-like hidden sector fermion considered in Ref.~\cite{Petriello:2008pu}.  This $U(1)_{\chi}$ $Z'$ state also acts as the messenger in the canonical Hidden Valley model of Ref.~\cite{Strassler:2006im}.  We emphasize, however, that the results of our study can be easily and straightforwardly generalized for decays to {\em any} hidden states using the formalism discussed in the next section.  We carefully consider the various systematic errors that affect this analysis.  We find that a 1 TeV $Z'$ can be discovered at the LHC with $50 \,{\rm fb}^{-1}$ of integrated luminosity, while 1.5 TeV states require super-LHC luminosities of $1\,{\rm ab}^{-1}$.  The discovery reach depends crucially on the systematic errors on the background, particularly the uncertainty on the dominant 
$Z\gamma$ background arising from imprecise knowledge of parton distribution functions (PDFs).

Our paper is organized as follows.  We present the signal process, its interpretation using effective charges, and our example $U(1)_{\chi}$ model in Section~\ref{sec:sig}.  In 
Section~\ref{sec:backs} we discuss the various backgrounds, estimate the uncertainty arising from imperfect knowledge of PDFs on the $Z\gamma$ background, and describe our analysis 
procedure.  We present our results for the LHC search reach in Section~\ref{sec:lhc}.  Finally, we conclude in Section~\ref{sec:conc}.

\section{Structure of signal process \label{sec:sig}}

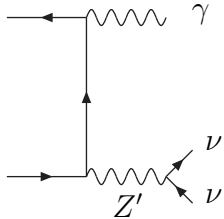
\begin{figure}[htb]
  \begin{center}
    \begin{picture}(100,100)(0,0)
      \SetColor{Black}
        \ArrowLine(0,10)(30,10)
        \ArrowLine(30,10)(30,70)
        \ArrowLine(30,70)(0,70)
        \Photon(30,70)(60,70){3}{4}
        \Photon(30,10)(60,10){3}{4}
        \put(70,70){$\gamma$}
        \put(40,-5){$Z'$}
        \ArrowLine(60,10)(70,20)
        \ArrowLine(70,0)(60,10)
        \put(75,0){$\nu$}
        \put(75,20){$\nu$}

    \end{picture}
  \end{center}
\caption{\label{IFdiags} 
          Example diagram giving rise to the signal process $pp \to \gamma Z' \to \nu\bar{\nu}+\met$.  The particle labeled $\nu$ can denote either a SM 
          neutrino or hidden sector state.}
\end{figure}

An example Feynman diagram leading to our signal process is shown in Fig.~(\ref{IFdiags}).  The cross section for this interaction can be written in the form 
\be
\sigma = \sigma_{ISR}^u + \sigma_{ISR}^d,
\ee  
where the superscripts $u,d$ denote contributions from initial-state up and down quarks and the subscript $ISR$ indicates the emission of the $Z'$ from the colliding 
particles.  We can write these two contributions in a form that makes it clear how they arise from the underlying charges of the $Z'$.  Each $\sigma_{ISR}^{u,d}$ can in turn be written as a product of two distinct terms: a piece which 
incorporates the matrix elements, parton distribution functions, and experimental cuts, denoted as $f_{ISR}^{u,d}$; a piece which depends on the charges 
from a given model, $Q_{ISR}^{u,d}$.  We then have $\sigma_{ISR}^{u,d} = f_{ISR}^{u,d}Q_{ISR}^{u,d}$.   For the process of Fig.~(\ref{IFdiags}), $ pp \rightarrow \gamma Z' \rightarrow \gamma X^\dagger X$, it can be shown that the couplings can be written in the form
\be
Q_{ISR}^q \equiv (q_V'^2+q_A'^2)\,Q_q^2\, \frac{\Gamma_{Z'}^{inv}}{\Gamma_{Z'}},
\ee
where $\Gamma_{Z'}^{inv}$ denote the partial widths of the $Z'$ to {\em any} invisible particle (SM $\nu$'s 
or hidden sector states), $\Gamma_{Z'}$ is the total width, and $Q_q$ is the electric charge of quark $q$ with the $\sqrt{4\pi\alpha}$ factor included.  
A prime on a charge indicates that it is a $Z'$ charge, while no prime denotes a SM charge.  The $A$ and $V$ subscripts denote axial and vector charges, respectively.   All model dependence is then encoded in $Q^{u,d}_{ISR}$, so that
any $Z'$ model can then be constructed by dialing $Q_{ISR}^{u,d}$ appropriately.  The functions $f_{ISR}^{u,d}$ depend on the given model 
under consideration only through the $Z'$ mass in the narrow width approximation.  As an example, we give below in Table~\ref{Qcharge} the numerical values for these couplings in the example 
model we use for illustration, a $U(1)_{\chi}$ model with an additional vector-like hidden state.    One need only rescale $Q_{ISR}^{u,d}$ for any given model of interest in order to compute the total production cross section, given $f_{u,d}$.  We also present the charges for a sequential $Z'$ model with an additional 
vector-like hidden particle.

The crucial fact that allows the LHC to search for the presence of hidden sectors is that the charges $Q_{ISR}^{u,d}$ are predicted if the $Z'$ boson decays invisibly only to neutrinos.  
They are known once the leptonic decays of the $Z'$ are measured in the Drell-Yan mode~\cite{Petriello:2008pu}.  Thus a prediction for the process of Fig.~(1) can be made from on-peak data, and an excess from $Z'$ decays to new states can be found.  Defining
\begin{eqnarray}
c_q&=&\frac{M_{Z'}}{24\pi\Gamma_{Z'}}(q_R^{\prime 2}+q_L^{\prime 2})(l_R^{\prime 2}+l_L^{\prime 2}), \nonumber \\ 
e_q&=&\frac{M_{Z'}}{24\pi\Gamma_{Z'}}(q_R^{\prime 2}-q_L^{\prime 2})(l_R^{\prime 2}-l_L^{\prime 2}), \nonumber \\ 
C&=&\frac{l_L^{\prime 2}}{l_R^{\prime 2}} = \frac{c_u-e_u-c_d+e_d}{c_u+e_u-c_d-e_d},
\end{eqnarray}
we can write 
\be
Q_{ISR}^q = \frac{c_q}{2} \frac{C}{C+1}  \frac{\Gamma_{Z'}^{inv}}{\Gamma_{Z'}^{\nu}}\,Q_q^2. 
\label{nupred}
\ee  
The quantities $c_q$, $e_q$ can be measured in Drell-Yan production~\cite{Petriello:2008pu}.  If the $Z'$ can decay invisibly only to neutrinos, 
then $\Gamma_{Z'}^{inv}/\Gamma_{Z'}^{\nu}=1$.  Any deviation from this prediction indicates the presence of additional hidden decays.

\begin{table}[htbp]
\centering
\begin{tabular}{|c|c|c|c|c|c|c|c|c|c|c|c|} \hline

& $Q_u^{ISR}$ & $Q_d^{ISR}$ & $u_L$ & $u_R$ & $d_L$ & $d_R$ & $e_L$ & $e_R$ & $X_L$ & $X_R$ \\ \hline
$U(1)_{\chi}^{hid}$ & 0.598 & 0.748 & $\frac{-1}{2\sqrt{6}}$ & $\frac{1}{2\sqrt{6}}$ & $\frac{-1}{2\sqrt{6}}$ & $\frac{-3}{2\sqrt{6}}$ & 
	$\frac{3}{2\sqrt{6}}$ & $\frac{1}{2\sqrt{6}}$ & 1 & 1 \\ \hline
$SSM^{hid}$ & 1.335 & 0.428 & $\frac{1}{2} - \frac{2}{3} s_W^2$ & $-\frac{2}{3} s_W^2$ & $-\frac{1}{2} + \frac{1}{3} s_W^2$ & 
$\frac{1}{3} s_W^2$ & $-\frac{1}{2} + s_W^2$ & $s_W^2$ & 1 & 1 \\ \hline 
	
\end{tabular}
\caption{\label{Qcharge} Numerical values of the $Q^{u,d}_{ISR}$'s for a $U(1)_{\chi}$ model and sequential $Z'$ with an additional hidden sector state $X$, multiplied by $10^3$.  We have also included the underlying charges of the considered model for orientation.  $s_W$ is the sine of the weak mixing angle.  In the sequential case, an overall factor of $g/c_W$ has been factored out, and is included in the hidden charges.  The mono-photon $Z'$ production cross-section can be computed for any model by re-scaling $Q^{u,d}_{ISR}$ for any model.  See the text for more details.}
\end{table}

\section{Backgrounds and analysis procedure \label{sec:backs}}	

Several distinct backgrounds can mimic the mono-photon signature of an invisibly decaying $Z'$:
\begin{enumerate}

\item the irreducible background $pp \to \gamma Z \to \gamma \nu \bar{\nu}$;
\item $pp \to \gamma W^{\pm} \to \gamma l^{\pm} \nu$, where the lepton (electron, muon or tau) is missed;
\item the Drell-Yan production process $pp \to W^{\pm}+X \to e^{\pm}\nu+X$ where the electron is misidentified as a photon and any additional jets are missed;
\item production of $Z+{\rm jets}$ where the Z decays invisibly and a jet fakes a photon;
\item high energy muons from cosmic rays or accelerator beam halo emitting bremsstrahlung photons while passing through electromagnetic calorimeter, giving rise to events with a reconstructed photon and missing transverse energy. 

\end{enumerate}
We impose the following preselection cuts on the photon candidate: $|\eta_{\gamma}|<1.5$ and $p_T^{\gamma}>100\,{\rm GeV}$.  For the $W\gamma$ background, 
we assume a 5\% possibility for a lepton to be missed for central rapidities; we also include contributions from all leptons outside the central region.  
For the Drell-Yan background, we assign a 2\% probability for an electron to 
fake a photon.  Both rates are consistent with Tevatron performance~\cite{tevrefs}.  In their mono-photon searches, the CDF and D0 collaborations obtain the $Z+{\rm jets}$ rate
as a fraction between $2-5\%$ of the irreducible $\gamma \nu \bar{\nu}$ background.  This value will likely be substantially larger at the LHC due to the increased importance of the gluon distribution function; while the $\gamma \nu \bar{\nu}$ is 
initiated at leading order via $q\bar{q}$ partonic interactions, $Z+{\rm jets}$ also has a $qg$ subprocess contribution.  We estimate this background in the following way.  
We use the preselection cuts discussed above and compute the ratio of the $Z+1\,{\rm jet}$ and $Z\gamma$ cross sections at both the Tevatron and the LHC, and use the 
increase in this ratio at the LHC to scale the measured Tevatron fake rate.  This leads to an estimate of the $Z+{\rm jets}$ background of $15-35\%$ of the 
$\gamma \nu \bar{\nu}$ rate.  We conservatively use the 35\% estimate.  For the Drell-Yan background we veto jets with $p_T>50\,{\rm GeV}$, which is a 
conservative estimate of LHC capabilities.  We use Madgraph~\cite{Maltoni:2002qb} to simulate both signal and background processes.  For the Drell-Yan background, we cross-check the result 
using PYTHIA~\cite{Sjostrand:2006za} to assure correct modeling of the electron $p_T$ spectrum.  We assume that the backgrounds from beam halo and 
cosmic rays contribute with a rate consistent with Tevatron findings.

\begin{figure}[htbp]
   \centering
   \includegraphics[width=0.7\textwidth,angle=0]{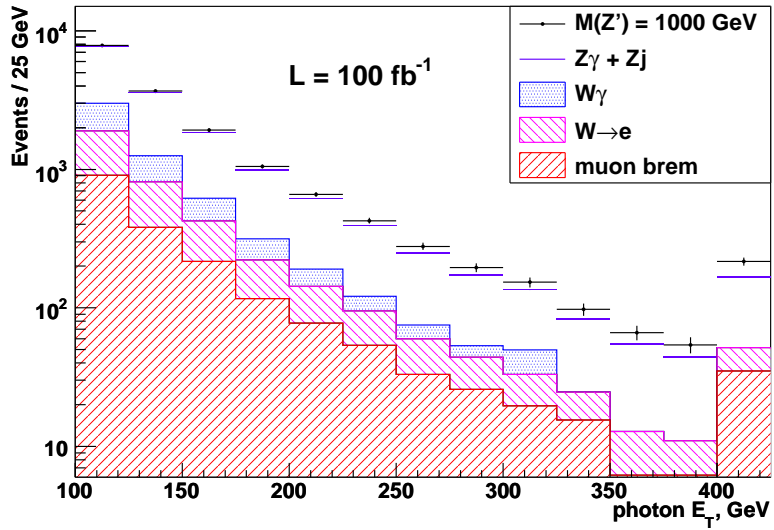}
   \caption{\label{crplot} Plot of signal and various backgrounds as a function of the photon $p_T$.  For the signal, a 1 TeV $U(1)_{\chi}$ model $Z'$ with 
   an additional hidden state has been assumed.  The last bin includes all photons with $p_T > 400$ GeV.  We note that the various histogrammed backgrounds are stacked atop each other.}
\end{figure}

We present in Table~\ref{crnums} the signal and background as a function of a lower $p_T^{\gamma}$ cut, using the parameters described above.  For the signal process 
we assume either a 1 or 1.5 TeV $U(1)_{\chi}$ $Z'$ boson.  Two kinematic handles separate the signal from the various backgrounds.  First, as is clear from the table, the $p_T^{\gamma}$ spectrum is harder for the signal than for any background process.  Second, the photon from the signal peaks at more central rapidities than the background.  This will be a useful 
experimental check, although we have not implemented any cut to exploit this in our analysis.  We present these results graphically in Fig.~(\ref{crplot}); the 
importance of understanding the $Z\gamma$ and $Zj$ backgrounds are viscerally clear from this plot.

\begin{table}[htbp]
\centering
\begin{tabular}{|c|c|c|c|c|c|c|c|}\hline
$p_T^{min} \,{\rm (GeV)}$ & $M_{Z'}=1\,{\rm TeV}$ & $M_{Z'}=1.5\,{\rm TeV}$ & $Z\gamma$ & $Z+{\rm jets}$ & DY $W$ & $W\gamma$ & Muon brem.\\ \hline\hline
100 & 11.8  &3.17 &135 & 47.3 & 34.6 & 34.7 & 18.9 \\ \hline
125 & 8.81 &2.46 &73.0 & 25.6 & 16.9 & 15.1 & 9.87\\ \hline
150 & 6.75 &1.94 &42.5 & 14.9 & 9.24 & 7.12 & 6.05\\ \hline
175 & 5.27 &1.57 &26.4 & 9.24 & 5.50 & 3.73 & 3.89 \\ \hline
200 & 4.18 &1.27 &17.5 & 6.13 & 3.63 & 2.05 & 2.73\\ \hline
225 & 3.35 &1.06 &11.9 & 4.17 & 2.45 & 1.20 & 1.95\\ \hline
250 & 2.72 &0.875 &8.33 & 2.92 & 1.71 & 0.738 & 1.41\\ \hline
275 & 2.22 &0.730 &6.02 & 2.11 & 1.23 & 0.462 & 1.08\\ \hline
300 & 1.83 &0.617 &4.43 & 1.55 & 0.903 & 0.297 & 0.826\\ \hline
325 & 1.51 &0.523 &3.30 & 1.16 & 0.664 & $-$ & 0.630\\ \hline
350 & 1.26 &0.443 &2.52 & 0.882 & 0.500 & $-$ & 0.475\\ \hline
375 & 1.06 &0.379 &1.96 & 0.686 & 0.382 & $-$ & 0.412\\ \hline
400 & 0.883 &0.323 &1.52 & 0.532 & 0.296 & $-$ & $-$\\ \hline

\end{tabular}
\caption{\label{crnums} Cross section results for the backgrounds and signal for $M_{Z'}=1$ and 1.5 TeV.  All results are in femto-barns.  Dashed entries indicate irrelevantly small rates.}
\end{table}

\subsection{PDF uncertainty study \label{sec:pdfs}}

We estimate here the expected error on the Standard Model prediction for $Z\gamma$ production at the LHC.  Although several kinematic 
features separate the $Z'$ signal from background, most notably the harder $p_T$ spectrum and shift of the photon to central rapidities in the $Z'$ case, it is clear from Table~\ref{crnums} 
that the $S/B$ ratio is such that normalization of the background is important.  A possible way to control the prediction for $Z\gamma$ production at the LHC is to normalize it to the 
cross section of $\gamma\gamma$, which proceeds through similar Feynman diagrams and partonic luminosities.  This involves writing the prediction for the $Z\gamma$ rate in the form
\begin{equation}
N_{Z\gamma} = \left(\frac{\sigma_{Z\gamma}}{\sigma_{\gamma\gamma}}\right)_{th} N_{\gamma\gamma}.
\label{zgnorm}
\end{equation}
The error on $N_{Z\gamma}$ is then driven by the statistical error on $N_{\gamma\gamma}$ and the uncertainty in the theory prediction for $\sigma_{Z\gamma}/\sigma_{\gamma\gamma}$.  
Since the $\gamma\gamma$ cross section is large and both PDF and scale uncertainties should cancel in the theoretical ratio, this should lead to a precise prediction.  This approach has been suggested in the literature to normalize di-boson predictions at the LHC to resonant $W,Z$ production~\cite{Dittmar:1997md}, and also to normalize missing energy plus jets to production of photons plus jets~\cite{newnorm}.

We study this by using 
the CTEQ 6.6 PDF fits~\cite{Nadolsky:2008zw} to compute the di-photon and $Z\gamma$ cross sections, their current PDF errors, and the PDF error on the  $\sigma_{Z\gamma}/\sigma_{\gamma\gamma}$ ratio.  Both the $Z\gamma$ and $\gamma\gamma$ processes are known at next-to-leading order in the 
QCD perturbative expansion, and progress toward next-to-next-to-leading order calculations of such di-boson processes is being made; we therefore 
expect the PDF uncertainty to ultimately be the limiting factor.  We require 
at least one photon have $|\eta|<1.5$ and allow the other to have  $|\eta|<2.5$ so that the di-photon cuts closely match those on $Z\gamma$, and study $p_T^{\min}=100,200$ GeV.  These leading 
order QCD results are presented below in Eq.~(\ref{pdfstud}):
\begin{eqnarray}
p_T>100\,{\rm GeV}&:& \sigma_{Z\gamma} =153^{+5}_{-5}\,{\rm fb},\,\,\, \sigma_{\gamma\gamma}= 595^{+23}_{-21}\,{\rm fb},\,\,\, \frac{\sigma_{Z\gamma}}{\sigma_{\gamma\gamma}} = 0.257^{+0.002}_{-0.003}; \nonumber \\ 
p_T>200\,{\rm GeV}&:&  \sigma_{Z\gamma} =19.2^{+0.8}_{-0.7}\,{\rm fb},\,\,\, \sigma_{\gamma\gamma}= 65.6^{+2.8}_{-2.3}\,{\rm fb},\,\,\, \frac{\sigma_{Z\gamma}}{\sigma_{\gamma\gamma}} = 0.292^{+0.003}_{-0.004}.
\label{pdfstud}
\end{eqnarray}
The branching fraction for $Z\to \nu\bar{\nu}$ has been included in these results.  It is clear that normalization to the di-photon cross section helps control the uncertainty in the prediction; the PDF error is reduced from 4\% to 1\% 
in the ratio, while the statistical error on $N_{\gamma\gamma}$ is roughly half of that on $N_{Z\gamma} $.  Comparing the uncertainty obtained by propagating through statistical and pdf errors 
directly on the $Z\gamma$ cross section to that found using Eq.~(\ref{zgnorm}) leads to an error reduction from 3.5\% to 1\% for $p_T^{min}=100$ GeV.

\section{LHC results \label{sec:lhc}}

We use our estimates of the signal and background $p_T^{\gamma}$ spectra to determine the required luminosity for both $3\sigma$ evidence and $5\sigma$ discovery at the LHC, using standard statistical tools~\cite{collie}.  The most crucial parameters in the analysis are the systematic errors on the various backgrounds.  We study here several estimates of the various efficiencies and systematic errors affecting this channel to determine their impact on the LHC discovery potential.  We first include an efficiency factor for reconstruction of the photon candidate in all signal and background processes of Table~\ref{crnums}.  We set this value to 56\%; this number is the average of the values found by the CDF and D0 experiments.  We incorporate the systematic errors listed in Table~\ref{syserrs} into our analysis.  We study two possible sets of systematic errors, which we deem ``very low'' and ``realistic.''  The efficiency error accounts for the uncertainty in the factor discussed above.  The K-factor/PDF systematic accounts for uncertainties arising from the QCD prediction for both the signal 
and $Z\gamma$ background.  The numbers used are motivated by the study in Section~\ref{sec:pdfs}.  The track veto error reflects the knowledge of how well the 2\% rate of electrons faking photons used in Section~\ref{sec:backs} can be determined.  The lepton veto error accounts for how well the 5\% probability for the lepton track to be missed in the $W\gamma$ background 
used in Section~\ref{sec:backs} will be known.

\begin{table}[htbp]
\centering
\begin{tabular}{|c|c|c|c|c|c|c|c|}\hline

 & Efficiency & K-factor/PDFs & Lepton veto & Track veto & Muon brem. \\ \hline\hline
 Very Low & 0.5 & 0.5 & 0.5 & 0.5 & 0.1 \\ \hline
 Realistic & 0.5 & 1.0 & 1.0 & 1.0 & 3.0 \\ \hline
 
\end{tabular}
\caption{\label{syserrs} Systematic errors affecting the various signal and background processes.  All numbers given are percent errors.  More detail regarding each is given in the text.}
\end{table}

\begin{figure}[htbp]
   \centering
   \includegraphics[width=0.7\textwidth,angle=0]{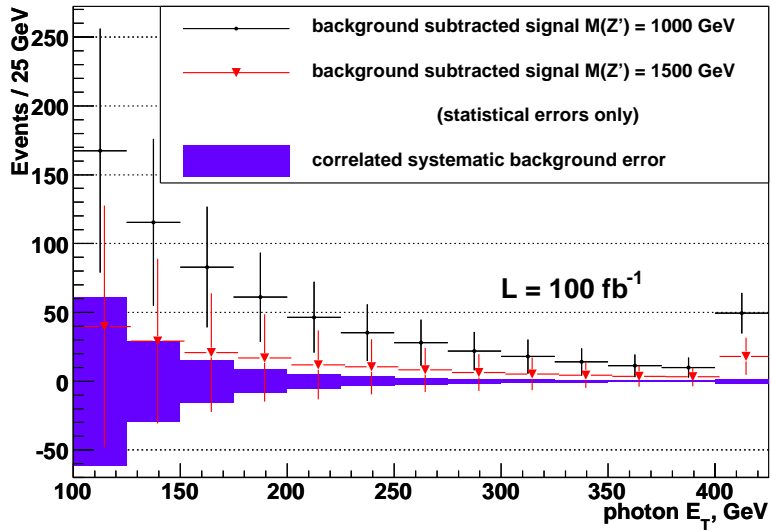}
   \caption{\label{errplot} Effects of "realistic" systematic and statistical errors in each $p_T^{\gamma}$ bin. Last bin includes overflows. The $U(1)_{\chi}$ model $Z'$ with 
   an additional hidden state has been assumed.}
\end{figure}

We present in Table~\ref{reqlum} the required integrated luminosity for $3\sigma$ evidence and $5\sigma$ discovery of both 1 and 1.5 TeV $U(1)_{\chi}$ Z' bosons at the LHC assuming the systematic errors  in Table~\ref{syserrs}.  For comparison we also show the results assuming only statistical errors.  Discovery of the 1 TeV invisibly decaying $Z'$ is possible with roughly $50\,{\rm fb}^{-1}$.  While the systematic errors degrade the search reach only slightly for 
$M_{Z'}=1\,{\rm TeV}$, they become crucial for heavier states.  A $5\sigma$ discovery of a 1.5 TeV state is possible only if systematic errors at the LHC can be controlled to the ``very low'' level; 
only $3\sigma$ evidence is possible for ``realistic'' errors.  The dominant systematic effect is the normalization of the $Z\gamma$ background.  Study of this background will be crucial to 
probe hidden sector decays of $Z'$ bosons in the mono-photon channel.  A plot showing the effects of both the systematic and statistical errors is presented in 
Fig.~(\ref{errplot}).

\begin{table}[htbp]
\centering
\begin{tabular}{|c|c|c||c|c|}\hline \multicolumn{5}{|c|}{SM $\nu$'s and New States}   \\ 
\multicolumn{3}{|c||}{$M_{Z'}=1\,{\rm TeV}$} & \multicolumn{2}{c|}{$M_{Z'}=1.5\,{\rm TeV}$}\\ \hline
 & $3\sigma$ & $5\sigma$ &  $3\sigma$ & $5\sigma$    \\ \hline\hline
None & 15 & 41 & 164 & 467 \\ \hline
Very Low & 16 & 43 & 216 & 1640  \\ \hline
Realistic & 18 & 54 & 795 & $-$ \\ \hline

\end{tabular}
\caption{\label{reqlum} Required integrated luminosity, in inverse femtobarns, to achieve both $3\sigma$ and $5\sigma$ signals at the LHC, for $Z'$ decay to new hidden states or SM neutrinos.  Included are results for the two mass points 
$M_{Z'}=1\,{\rm TeV}$ and $M_{Z'}=1.5\,{\rm TeV}$.  The dash indicates that $5\sigma$ discovery of the 1.5 TeV state is not possible with ``realistic'' errors.}
\end{table}

\begin{figure}[htbp]
   \centering
   \includegraphics[width=0.6\textwidth,angle=90]{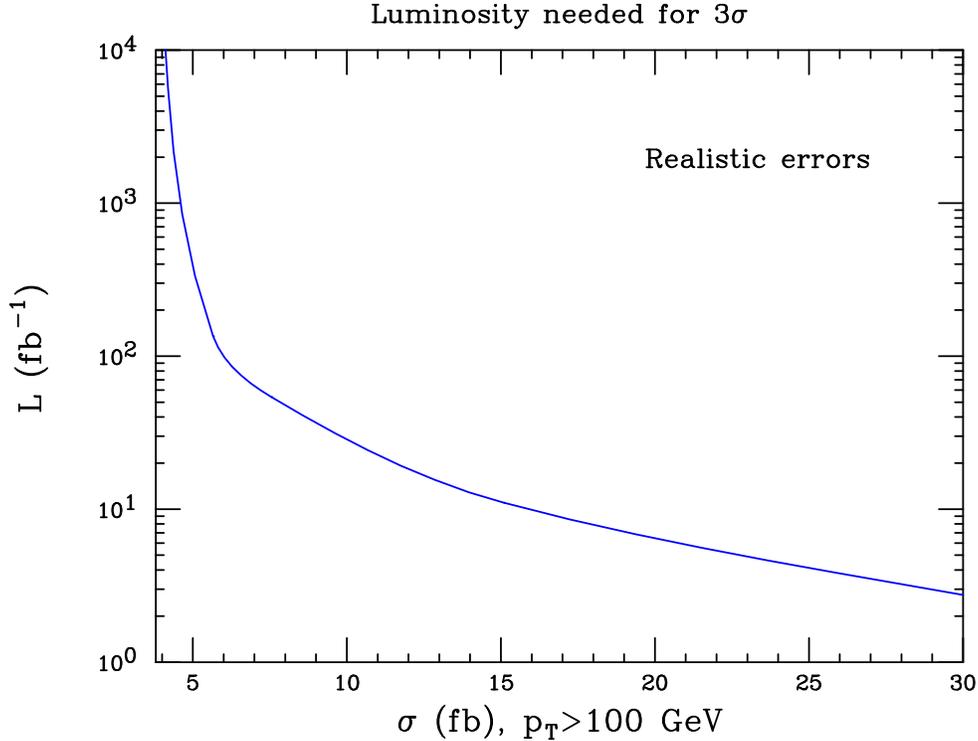}
   \caption{\label{pppplot} Required integrated luminosity for $3\sigma$ evidence of hidden sector $Z'$ decays for a given cross section $\sigma$ into hidden states, including 
   	the cut $p_T^{\gamma}>100$ GeV.  We have assumed the realistic errors described in the text.}
\end{figure}

The numbers in Table~\ref{reqlum}, however, only provide the discovery of $Z'$ decays to any invisible state, whether SM neutrino or hidden sector state.  We wish to determine whether decays of the $Z'$ to new states can be separated from decays to SM neutrinos.   To analyze whether this is possible, we add decays to SM neutrinos to the background, and check if an excess of the rate over that predicted by on-peak data can be observed.  This is possible since Eq.~(\ref{nupred}) predicts the production cross section for $pp \rightarrow Z' \gamma \rightarrow \bar{\nu} \nu \gamma$ once 
$pp \rightarrow Z' \rightarrow \ell^+ \ell^-$ is measured.  This relies only upon the mild assumption that the charges do not break $SU(2)$ invariance. The required luminosity to rule out $Z'$ decays to hidden sectors at 95\% C.L. is shown in Table~\ref{HidOnly}.   For ``realistic'' errors, a 10\% error on the cross-section for $pp \rightarrow Z' \rightarrow \ell^+ \ell^-$ has been added, to account for the uncertainty on the prediction for $pp \rightarrow Z' \rightarrow \bar{\nu} \nu$.  We also show in Fig.~(\ref{pppplot}) the size of the production cross-section, for $pp \rightarrow Z' \gamma \rightarrow X^\dagger X \gamma$, where $X$ is a new hidden state, which can be observed at $3 \sigma$ as a function of the luminosity.

A previous study of the  $pp \to ZZ' \to \ell^+ \ell^- \met$ mode found that a $5\sigma$ discovery of an invisibly decaying $Z'$ was possible with slightly over 
$30\,{\rm fb}^{-1}$, while $3\sigma$ evidences requires roughly $15\,{\rm fb}^{-1}$~\cite{Petriello:2008pu}.  The required integrated luminosities for the $ZZ'$ mode are similar to the values found here.  
Although the study in Ref.~\cite{Petriello:2008pu} did 
not include systematic errors, they are expected to be smaller for the $ZZ'$ channel, and since the $S/B$ is larger in that mode finding the invisible $Z'$ should be less sensitive to such effects.  Discovery and study of an invisibly decaying $Z'$ are possible in both channels.
\begin{table}[htbp]
\centering
\begin{tabular}{|c|c|c|c||c|c|c|}\hline \multicolumn{7}{|c|}{New States Only}   \\ 
\multicolumn{4}{|c||}{$M_{Z'}=1\,{\rm TeV}$} & \multicolumn{3}{c|}{$M_{Z'}=1.5\,{\rm TeV}$}\\ \hline
 & $N$ & $0.5\times N$ & $2 \times N$ & $N$ & $0.5\times N$ & $2 \times N$     \\ \hline\hline
None & 20 & 79 & 5.3 & 203&807&51 \\ \hline
Very Low & 22&105& 5.5 & 372 & $-$ & 62  \\ \hline
Realistic & 24 & 171 & 5.6 & $-$ & $-$ & 72\\ \hline

\end{tabular}
\caption{\label{HidOnly} Required integrated luminosity, in inverse femtobarns, to rule out at 95\% C.L. $Z'$ decays to any non-Standard Model state.  $N$ is for the ``nominal'' hidden sector of Table~2, and $0.5 \times N$ and $2 \times N$ denote hidden sectors with half and twice the nominal production cross sections, respectively. Included are results for the two mass points 
$M_{Z'}=1\,{\rm TeV}$ and $M_{Z'}=1.5\,{\rm TeV}$.   In the nominal model of  Table~2, decays to hidden sector states constitute approximately 2/3 of invisible $Z'$ decays.}
\end{table}

\section{Conclusions \label{sec:conc}}

In this paper we have studied whether invisible decays of $Z'$ bosons to light hidden particles can be discovered at the LHC using the channel $pp \to \gamma Z' \to \gamma \met$.  This process 
occurs when the $Z'$ decays to new hidden states, such as appear in Hidden Valleys, some extensions of the MSSM, extra dimensions, and in many models of dark matter.
We show that 
this process can be simply described using only the $Z'$ mass and two effective charges.  If the $Z'$ decays invisibly only to Standard Model neutrinos, then these charges are predicted by observation 
of the $Z'$ through the Drell-Yan process, allowing for the separation of invisible decays to SM neutrinos from new hidden states.  We enumerate the various backgrounds that lead to the mono-photon signature, and estimate the systematic errors on the background rates at the LHC.  We find that with $50\,{\rm fb}^{-1}$ of integrated luminosity, a 1 TeV $Z'$ state can be discovered with $5\sigma$ significance.  $Z'$ bosons with 1.5 TeV masses require super-LHC 
luminosities of $1\,{\rm ab}^{-1}$.  Control of the systematic errors, in particular the normalization of the dominant $Z\gamma$ Standard Model background, is crucial to maximize the LHC search 
reach.

Discovery of a hidden sector, or new hidden states, would be an exciting advance in our understanding of Nature.  We have shown that it is feasible using the mono-photon channel at the LHC.  The possibility of observing such states through the hidden decays of a new vector gauge boson makes the accurate measurement of invisible $Z'$ decays at the LHC an exciting and 
reachable goal.

\section*{Acknowledgments}

The authors are supported by the DOE grants DE-FG02-95ER40896 and DE-FG02-97ER41022, 
Outstanding  Junior Investigator Awards, 
by the University of Wisconsin Research Committee
with funds provided by the Wisconsin Alumni Research Foundation, and
by the Alfred P.~Sloan Foundation.

\end{document}